\documentclass[aps,titlepage,preprint]{revtex4-2}
\usepackage[utf8]{inputenc}
    \usepackage[margin=1in]{geometry}
    \usepackage{amsmath}
\usepackage{xcolor}
\usepackage{braket}
\usepackage{mathrsfs}
\usepackage{amsfonts}
\usepackage{epsfig,float}
\usepackage[normalem]{ulem}

\usepackage{amssymb,amsfonts}

\newcommand{\mc}{\mathcal}

\newcommand{\ms}{\mathscr}

\newcommand{\bb}{\mathbb}

\newcommand{\defn}{\mathrel\equiv} 

\newcommand{\scri}{\ms I}

\definecolor{indigo(dye)}{rgb}{0.0, 0.25, 0.42}

\usepackage[colorlinks, allcolors=blue!70!black, linktocpage]{hyperref}

\usepackage{cleveref}

\crefname{equation}{Eq.}{Eqs.}
\creflabelformat{equation}{(#2#1#3)}

\crefname{section}{\S}{\S}
\crefname{appendix}{Appendix}{Appendices}
\crefname{figure}{Fig.}{Figs.}

\crefname{definition}{Def.}{Defs.}
\crefname{prop}{Prop.}{Props.}
\crefname{lemma}{Lemma}{Lemmas}
\crefname{corollary}{Cor.}{Cors.}
\crefname{thm}{Theorem}{Theorems}
\crefname{remark}{Remark}{Remarks}

\crefname{ass}{Assumptions}{Assumptions}
\crefname{property}{Properties}{Properties}

\begin{document}
\title{Black Holes Decohere Quantum Superpositions}
\author{Daine L. Danielson}\email{Corresponding author: daine@uchicago.edu} \author{Gautam Satishchandran}\email{gautamsatish@uchicago.edu}
\author{Robert M. Wald}\email{rmwa@uchicago.edu}
\affiliation{Enrico Fermi Institute and Department of Physics, \\The University of Chicago,  Chicago, Illinois 60637, USA}
\date{\today}

\begin{abstract}
\noindent We show that if a massive body is put in a quantum superposition of spatially separated states, the mere presence of a black hole in the vicinity of the body will eventually destroy the coherence of the superposition. This occurs because, in effect, the gravitational field of the body radiates soft gravitons into the black hole, allowing the black hole to acquire ``which path'' information about the superposition. A similar effect occurs for quantum superpositions of electrically charged bodies. We provide estimates of the decoherence time for such quantum superpositions. We believe that the fact that a black hole will eventually decohere any quantum superposition may be of fundamental significance for our understanding of the nature of black holes in a quantum theory of gravity.

\vspace{24pt}
{\em This essay is awarded third prize in the 2022 Essay Competition of the Gravity Research Foundation.}
\end{abstract}
\maketitle

Black holes have long been known to be destroyers of quantum coherence. If one member of an entangled pair of particles falls into a black hole, all that will remain is the particle that stayed outside the black hole, which will be in a mixed state. Much more generally, if matter of any kind falls into a black hole, it will, in effect, eventually emerge as Hawking radiation and be in a highly mixed state. While it may be debated as to whether the quantum coherence is lost forever in this process (see, e.g., \cite{Unruh_2017,Marolf_2017}), there is a general consensus that the state outside the black hole is highly mixed at least up to the ``Page time'' in black hole evaporation.

The purpose of this essay is to show that black holes are even more insidious destroyers of quantum coherence than has been previously known. If one puts any quantum matter in a spatial superposition, the mere presence of a black hole in the vicinity of the matter will eventually destroy the coherence of this superposition. This happens because the long-range (i.e., electromagnetic and gravitational) fields associated with the quantum matter affect the quantum state of these fields on the black hole horizon. In effect, the black hole thereby acquires ``which path'' information about the quantum superposition. As we shall show, this inflicts a fundamental rate of decoherence even on stationary superpositions outside its event horizon. This is sufficient to decohere any quantum superposition over a sufficiently long period of time.

To understand how this works, it is useful to first consider a quantum superposition in flat spacetime and see how decoherence can be avoided, following the analysis given in Ref. \cite{Danielson_2021}. For simplicity and definiteness, we consider an electrically charged body and the decoherence effects of the electromagnetic field, but an exactly similar analysis will apply for a massive body in the gravitational case. Below, we will refer to the charged body as a ``particle'' although it need not be an elementary particle, e.g. it could be an atom or a nanoparticle. Suppose an experimenter, Alice, sends a particle of charge $q$ with spin initially in the positive $x$-direction through a Stern-Gerlach apparatus oriented in the $z$-direction, so that the state of her particle after the process is in a superposition state of the following form: 
\begin{equation}
\label{eq:Alicepart}
\frac{1}{\sqrt{2}}\big(\ket{\uparrow;\mc{A}_{1}}+\ket{\downarrow;\mc{A}_{2}}\big) \, .
\end{equation}
Here $\ket{\mc{A}_{1}}$ and $\ket{\mc{A}_{2}}$ are spatially separated wavepackets with separation $d$, with $\ket{\uparrow}$ and $\ket{\downarrow}$ being eigenstates of the $z$-spin. We wish to know whether the coherence of this superposition is preserved at a later time. In order to make this into a  well defined experimental/observational question, Alice can put the particle through a reversing Stern-Gerlach apparatus at some later time and measure the $x$-spin. If the coherence of the superposition \cref{eq:Alicepart} has been maintained, the spin will always be found to be in the positive $x$-direction, whereas if any coherence has been lost the spin will sometimes be found to be in the negative $x$-direction. 

We assume that there are no external influences whatsoever on Alice's particle. It might then seem obvious that coherence must be maintained. However, this is not necessarily the case because, since the particle is charged, an electromagnetic field is present and it is part of the system. Heuristically, the state of the total system after passage through the initial Stern-Gerlach apparatus is actually of the following form:
\begin{equation}
\label{eq:Aliceinst2}
\frac{1}{\sqrt{2}}\big(\ket{\uparrow;\mc{A}_{1}} \otimes \ket{\psi_1}+\ket{\downarrow;\mc{A}_{2}} \otimes \ket{\psi_2}\big)
\end{equation}
where $\ket{\psi_{1}}$ and $\ket{\psi_{2}}$ formally correspond to the states of the electromagnetic field for the charge-current sources determined by $\ket{\mc{A}_1}$ and $\ket{\mc{A}_2}$, respectively. Since $\ket{\psi_{1}}$ and $\ket{\psi_{2}}$ clearly are distinguishable electromagnetic fields, it might seem that Alice's particle is already decohered at the outset. However, this decoherence is a ``false decoherence'' in the sense of Ref. \cite{unruh_2011}. If Alice recombines her particle slowly enough so as to avoid radiating, she will be able to fully restore the coherence of her particle.

In order to give a precise description of the true decoherence of Alice's particle associated with the electromagnetic field, it is necessary to separate the electromagnetic field into a ``Coulomb part'' (which is not an independent degree of freedom and should cause only a false decoherence of Alice's particle) and a ``radiation part'' (which corresponds to the true degrees of freedom of the electromagnetic field that should be responsible for a true decoherence, observable by Alice). In general, this distinction is not possible to make in a meaningful way at any finite time. However, the situation improves considerably if we go to asymptotically late times. At asymptotically late times, the electromagnetic field naturally decomposes into a radiation field that propagates to null infinity and a Coulomb field that follows Alice's particle to timelike infinity. The asymptotic Coulomb field is completely determined by the asymptotic state of Alice's particle and does not represent an independent degree of freedom. Thus, at asymptotically late times, the state of the total system is of the following form:
\begin{equation} 
\label{eq:Aliceinst3}
\frac{1}{\sqrt{2}}\big(\ket{\uparrow;\mc{A}_{1}}_{i^+} \otimes \ket{\Psi_1}_{\scri^{+}}+\ket{\downarrow;\mc{A}_{2}}_{i^+}  \otimes \ket{\Psi_2}_{\scri^{+}}\big).
\end{equation} 
Here $\ket{\uparrow;\mc{A}_{1}}_{i^+} $ and $\ket{\downarrow;\mc{A}_{2}}_{i^+} $ represent the asymptotically late-time states of the components of Alice's particle and $\ket{\Psi_1}_{\scri^{+}}$ and $\ket{\Psi_2}_{\scri^{+}}$ represent the quantum states of the electromagnetic radiation at future null infinity $\ms{I}^{+}$. If Alice has recombined her particle at some finite time, then $\ket{\mc{A}_{1}} = \ket{\mc{A}_{2}}$. Thus, the decoherence of Alice's superposition will be determined by the orthogonality of the radiation states
\begin{equation}
\label{decAlicescri}
{\mathscr D} = 1 - \left\lvert\braket{\Psi_{1}|\Psi_{2}}_{\scri^{+}}\right\rvert.
\end{equation}
In the absence of any external influences, Alice can ensure that the coherence of her particle is maintained (i.e. ${\mathscr D} \approx 0$) if she recombines her particle in such a way that negligible entangling radiation is emitted. As estimated in Ref. \cite{Belenchia_2018}, this will be possible if the recombination is done over a time span $T$ such that
\begin{equation}
T \gg \frac{qd}{\sqrt{\epsilon_{0}c^{3}\hbar}} .
\label{timespan}
\end{equation}
In other words, if \cref{timespan} holds, Alice can ensure that $\ket{\Psi_1}_{\scri^{+}} \approx \ket{\Psi_2}_{\scri^{+}} \approx \ket{0}_{\scri^{+}}$, so ${\mathscr D} \approx 0$.
Thus, in Minkowski spacetime, Alice can, in principle, maintain the quantum coherence of her spatial superposition by recombining the components of the superposition slowly enough.

We now consider how this situation changes if there is a black hole in the vicinity of Alice. First, Alice must ensure that her lab does not fall into the black hole. One way of doing this would be for Alice to orbit the black hole. However, this may result in some unwanted emission of radiation. Therefore it would be better to equip Alice with a rocket engine that keeps her lab stationary. She must then also apply some force to her particle (e.g., via a uniform electric field) to keep it stationary. There also may be other effects in her lab due to the spacetime curvature associated with the black hole. However, Alice can take the effects of the gravitational field of the black hole on her lab into account in such a way that they will not will not result in the decoherence of her particle. Therefore, we shall ignore these effects. However, as we shall now explain, the black hole itself will acquire ``which path'' information about Alice's particle, which will result in decoherence.

With regard to the decoherence of Alice's particle, the key difference arising when a black hole is present is that electromagnetic radiation can now propagate through the black hole horizon as well as to null infinity. Thus, when a black hole is present, the asymptotically late-time state of Alice's particle and the electromagnetic field is now
\begin{equation} 
\label{eq:Aliceinst3}
\frac{1}{\sqrt{2}}\big(\ket{\uparrow;\mc{A}_{1}}_{i^+} \otimes \ket{\Psi_1}_{\scri^{+}}\ket{\Phi_1}_{\ms{H}^{+}}+\ket{\downarrow;\mc{A}_{2}}_{i^+}  \otimes \ket{\Psi_2}_{\scri^{+}} \ket{\Phi_2}_{\ms{H}^{+}} \big)
\end{equation} 
where $\ket{\Psi_1}_{\scri^{+}}$ and $\ket{\Psi_2}_{\scri^{+}}$ are as before and $\ket{\Phi_1}_{\ms{H}^{+}}$ and $\ket{\Phi_2}_{\ms{H}^{+}}$ are the corresponding states of the electromagnetic field on the event horizon, $\ms{H}^{+}$, of the black hole. The decoherence of Alice's particle in the presence of a black hole is now given by
\begin{equation}
\label{decAlicescribh}
{\mathscr D} = 1 - \left\lvert\braket{\Psi_{1}|\Psi_{2}}_{\scri^{+}} \braket{\Phi_{1}|\Phi_{2}}_{\ms{H}^{+}}\right\rvert.
\end{equation}
As in Minkowski spacetime, if Alice recombines her particle adiabatically, she can ensure that there is negligible radiation to infinity, so $\ket{\Psi_1}_{\scri^{+}} \approx \ket{\Psi_2}_{\scri^{+}} \approx \ket{0}_{\scri^{+}}$, in which case any decoherence will be entirely due to radiation propagating into the black hole
\begin{equation}
\label{decAlicescribh}
{\mathscr D}_{\rm BH} = 1 - \left\lvert \braket{\Phi_{1}|\Phi_{2}}_{\ms{H}^{+}}\right\rvert.
\end{equation}
It might be thought that, by performing her recombination adiabatically, Alice also can ensure that no radiation enters the black hole. However, this is not the case.

To see this, we first consider a classical point charge outside of a Schwarzschild black hole. The explicit solution for a static point charge outside of a Schwarzschild black hole has long been known \cite{Copson_1928,Cohen_1971,Linet_1976}. On the horizon, the electric field of a static point charge is purely radial, i.e. the only nonvanishing component of the electric field on the horizon is $E_{r}= c F_{ab}\ell^{a}n^{b}$, where $n^a$ denotes the affinely parametrized null normal to the horizon and $\ell^a$ is the unique past-directed radial null vector satisfying $\ell^a n_a = 1$. Electromagnetic radiation on the horizon is described by the pullback, $E_A$, of the electric field $E_a = c F_{ab} n^{b}$ to the horizon, where capital Latin indices denote angular components on the horizon. Since $E_A = 0$ for a static point charge, there is no radiation through the horizon, as would be expected. However, suppose we now quasi-statically move the point charge to a new location. After the charge has reached its new location, the electric field will again be radial, but $E_{r}$ on the horizon will be different from what it was initially. However, it follows from Maxwell's equations at the horizon that
\begin{equation}
\label{eq:MWeqs}
    \mc{D}^{A}E_{A}=-\partial_{V}E_{r}
\end{equation}
where $\mc{D}_{A}$ denotes the covariant derivative on the $2$-sphere cross-sections of the horizon, angular indices are raised and lowered with the metric, $q_{AB}$, on the cross-section, and $V$ is an affine parameter such that $n^a = (\partial/\partial V)^a$. Therefore, we must have $E_{A} \neq 0$ on the horizon as the charge is being moved and, indeed, $\int E_A dV$ 
is constrained by initial and final values of $E_{r}$, independently of how the charge is moved between its initial and final positions. Thus, there is necessarily some radiation that crosses the horizon of the black hole due to the displacement of the charge.
We can make the total energy flux of this radiation through the horizon arbitrarily small by moving the charge very slowly, but, as we will now show, we cannot make the ``total photon flux'' of this radiation small by moving the charge quasi-statically.

In order to analyze quantum aspects of the radiation, we need to give a precise specification of the quantum state of electromagnetic radiation on the horizon of a black hole. For an unperturbed Schwarzschild black hole formed by gravitational collapse, the state of the electromagnetic field on the horizon of the black hole is described by the Unruh vacuum. However, we will be concerned here only with low frequency phenomena ($\omega \ll 1$), in which case the Unruh and Hartle-Hawking vacua near the horizon are essentially indistinguishable. For the electromagnetic field in a gauge where $A_a n^a = 0$ on the horizon, the ``free data'' of the electromagnetic field on the horizon is the pull-back, $A_{A}$, of the vector potential. In the Fock space associated with the Hartle-Hawking vacuum, a ``particle'' corresponds to a solution that is purely positive frequency with respect to affine parameter on the horizon. The inner product on the one-particle Hilbert space is given by \cite{Kay:1988mu}
\begin{equation}
\label{eq:innerprod}
    \braket{A_{1,B}|A_{2,C}}_{\ms{H}^{+}}\defn \frac{2\epsilon_{0} c}{\hbar } \int_{\bb{S}^{2}} r_{\text{s}}^2 d\Omega \int_{0}^{\infty}\frac{\omega d\omega}{2\pi} ~q^{BC}\overline{{\hat{A}_{1,B}}(\omega,x^{A})}{\hat{A}_{2,C}}(\omega,x^{A})
\end{equation}
where $r_{\text{s}}$ is the Schwarzschild radius of the black hole and $\hat{A}_{A}$ is the Fourier transform of $A_{A}$ with respect to affine parameter $V$. \Cref{eq:innerprod} corresponds to a Klein-Gordon type of inner product on the positive frequency part of $A_{A}$. Now suppose that the black hole is perturbed by a classical charge-current source of the quantum electromagnetic field. The quantum state of the electromagnetic field will then be described by the coherent state (relative to the unperturbed vacuum) associated with the classical retarded solution. The expected number of ``horizon photons'' in this electromagnetic state at the horizon is given by
\begin{equation}
\label{eq:horphoton}
\langle N \rangle =  \Vert A_{A}\Vert^2_{\ms{H}^{+}},
\end{equation}
where $A_A$ is the classical retarded solution and the norm of $A_{A}$ is defined by the inner product \cref{eq:innerprod}.

Let us apply this result to the electromagnetic field of a point charge that starts at a point $x$ outside the black hole, is moved to another point $x'$ outside of the black hole and remains at $x'$ forever. We have already seen in this case that $\int E_A dV \neq 0$. Since $E_A = -c\partial_V A_A$, this means that $A_A$ does not return to its initial value at the end of the process. This is closely analogous to the memory effect that occurs at null infinity \cite{Bieri_2013,Satishchandran_2019}. The fact that $A_A$ does not return to its initial value implies that its Fourier transform diverges as $1/\omega$ as $\omega \to 0$. It then follows immediately from \cref{eq:innerprod} that $\Vert A_A\Vert^2_{\ms{H}^{+}} = \infty$. Thus, if one moves a point charge from $x$ to $x'$ and leaves the particle at $x'$ forever, no matter how quickly or slowly the charge is moved, an infinite number of ``soft horizon photons'' will be radiated into the black hole. This is closely analogous to the infrared divergences at null infinity that arise in scattering theory in quantum electrodynamics \cite{Satishchandran_2021}. Note that the infinite number of ``soft photons'' carry negligible energy, and by moving the charge quasi-statically, the total electromagnetic energy radiated into the black hole can be made to be arbitrarily small.

The case of more relevance for us is one in which the point charge is moved from $x$ to $x'$, is held at $x'$ for a long proper time $T$, and then is returned to $x$. In that case, $A_A$ returns to its initial value at late times, so there is no infrared divergence in the sense that $\langle N \rangle$ is finite. Nevertheless, the following estimates show that $\langle N \rangle$ is very large when $T$ is very large. The radial electric field of a point charge located a distance $b$ from the black hole is roughly $E_{r} \sim q/\epsilon_{0}b^{2}$ \cite{Copson_1928,Cohen_1971,Linet_1976}. The change in the radial electric field when the charge is moved from $x$ to $x'$ is therefore roughly $\Delta E_{r} \sim qd/\epsilon_{0}b^{3}$, where $d$ is the distance between $x$ and $x'$ and we have assumed that $d \ll b$. Taking account of the fact that the $2$-spheres on the horizon are of radius $2 G M/c^2$, it then follows from \cref{eq:MWeqs} that the change in the vector potential, $A_A$, on the horizon when the particle is moved from $x$ to $x'$ is
\begin{equation}
\label{eq:Achange}
\Delta A_{A} \sim \frac{G^2 M^2}{c^{5}} \frac{qd}{\epsilon_{0}b^{3}}.
\end{equation}
Eventually, when the particle is moved back to $x$, the change in $A_A$ will be equal and opposite to \cref{eq:Achange}. But if the charge is held at point $x'$ for a very long time $T$, the contribution of $A_{A}$ to the norm defined by \cref{eq:innerprod} will be dominated by the low-frequency contribution arising from the time interval over which \cref{eq:Achange} holds. We obtain
\begin{equation}
\label{eq:AliceNorm}
 \langle N \rangle = \Vert A_{A}\Vert_{\ms{H}^{+}}^{2}\sim\frac{G^{4}M^{4} q^2 d^2}{\hbar c^{9}\epsilon_{0}b^{6}}\ln V
\end{equation}
where $V$ is the affine time on the horizon corresponding to the proper time $T$ along the particle trajectory. However, the relation between affine time, $V$, and Killing time, $v$, on the horizon of a black hole is given by $V = \exp(\kappa v/c)$, where $\kappa=c^{4}/4GM$ is the surface gravity of the black hole. Furthermore, the Killing time is related to the proper time of the particle by the redshift factor. We shall assume that Alice's lab is not extremely close to the black hole and neglect the departure of the redshift factor from $1$. We then obtain
\begin{equation}
\label{eq:AliceNorm}
 \langle N \rangle = \Vert A_{A}\Vert_{\ms{H}^{+}}^{2}\sim \frac{G^{3}M^{3} q^2 d^2}{\hbar c^{6}\epsilon_{0}b^{6}}T.
\end{equation}
Thus, the number of ``soft photons'' radiated into the black hole in the above process grows linearly with the time, $T$, that the point charge spends at point $x'$.

We now have all of the ingredients needed to analyze Alice's coherence experiment, under the assumption that Alice splits and recombines her particle slowly enough that negligible radiation is emitted to infinity. Although our results hold much more generally, it is easiest to consider the case where, after passing through the Stern-Gerlach apparatus, the first component of Alice's particle remains at position $x$ and the second component of her particle moves to position $x'$. After these components stay at $x$ and $x'$, respectively, for a time $T$, they are recombined in such a way that the first component continues to remain at $x$ and the second component moves from $x'$ to $x$. In that case, no radiation is emitted by the first component, so in \cref{eq:Aliceinst3}, we have $\ket{\Phi_1}_{\ms{H}^{+}} = \ket{0}_{\ms{H}^{+}}$. However, our above analysis applies to the second component, which moves from $x$ to $x'$, stays at $x'$ for a time $T$, and then returns to $x$. Thus, $\ket{\Phi_2}_{\ms{H}^{+}}$ will be a state with expected number of photons given by \cref{eq:AliceNorm}. If $\langle N \rangle \gtrsim 1$, then $\ket{\Phi_2}_{\ms{H}^{+}}$ will be nearly orthogonal to $\ket{\Phi_1}_{\ms{H}^{+}} = \ket{0}_{\ms{H}^{+}}$. This means that---due entirely to the presence of a black hole---Alice's particle will decohere in a time
\begin{align}
    T_D &\sim \frac{\epsilon_{0}\hbar c^{6}b^{6}}{G^{3}M^{3} q^2 d^2} \\
        & \sim 10^{43}~\text{years}~ \bigg(\frac{b}{\text{a.u.}}\bigg)^{6}\cdot \bigg(\frac{M_\odot}{M}\bigg)^{3}\cdot \bigg(\frac{e}{q}\bigg)^{2} \cdot \bigg(\frac{\rm{m}}{d}\bigg)^{2}.
\end{align}
Thus, if our Sun were a black hole and if one separated an electron into two components one meter apart in a laboratory experiment on Earth, it would not be possible to maintain the coherence of the electron for more than $10^{43}$ years. On the other hand, if this experiment were done at $b= 6GM/c^2$, then $T_D \sim 5 \, {\rm minutes}$.

A closely parallel analysis can be given for the case of a gravitating particle. In the gravitational case, the electric part of the Weyl tensor $E_{ab}= c^2 C_{acbd}n^{c}n^{d}$ plays a role closely analogous to the role played by the electric field $E_a$ in the electromagnetic case. For a static point mass outside a Schwarzschild black hole the only non-vanishing component of the electric part of the Weyl tensor on the horizon is $E_{rr}=c^2 C_{acbd}\ell^{a}n^{c}\ell^{b}n^{d}$. Gravitational radiation on the horizon is described by the pullback, $E_{AB}$, of $E_{ab}$, which vanishes for a static point mass. However, the process of moving the particle quasi-statically to a new location will involve a change in $E_{rr}$. The (once-contracted) Bianchi identity on the horizon yields
\begin{equation}
\label{eq:EABErr1}
    \mc{D}^{A}E_{AB}=-\partial_{V}E_{rB}, \quad  \quad \mc{D}^{A}E_{rA}=-\partial_{V}E_{rr}
\end{equation}
which implies
\begin{equation}
\mc{D}^{A}\mc{D}^{B}E_{AB} = \partial_{V}^{2}E_{rr}
\end{equation}
in close analogy with \cref{eq:MWeqs}. Thus, if a point mass is moved quasi-statically, there necessarily will be radiation through the horizon. To determine the number of gravitons emitted, we treat the quantum gravitational field at the level of linearized perturbation theory about the black hole background. For a metric perturbation $h_{ab}$ in a gauge where $h_{ab}n^{a}=0=q^{AB}h_{AB}$ on the horizon, the ``free data'' on the horizon is $h_{AB}$. As in the electromagnetic case, a ``particle'' in the Fock space associated to the Hartle-Hawking vacuum is a solution with positive frequency with respect to affine parameter $V$. The inner product on the one-particle Hilbert space is given by a direct analog of \cref{eq:innerprod} with $A_A$ replaced by $h_{AB}$. Finally, $E_{AB}$ is given in terms of $h_{AB}$ by $E_{AB} = -\tfrac{1}{2}c^{2}\partial^2_V h_{AB}$.

The analysis of the decoherence of a quantum superposition of a body of mass $m$ in the presence of a black hole now proceeds in exact parallel with the electromagnetic case. The only significant difference is that, for the same reason as in the analysis of Ref. \cite{Belenchia_2018}, it is now the effective mass quadrupole $m d^2$ of the superposition that enters, rather than the effective electrostatic dipole $qd$ that entered the electromagnetic analysis. We find that a black hole will decohere a quantum superposition of a massive body in a time
\begin{align}
    T_D^{\text{GR}} &\sim \frac{\hbar  c^{10}b^{10}}{G^{6}M^{5}m^2 d^4} \\
    & \sim 10\ \mu\text{s}~ \bigg(\frac{b}{\text{a.u.}}\bigg)^{10}\cdot \bigg(\frac{\text{M}_\odot}{M}\bigg)^{5}\cdot \bigg(\frac{\text{M}_{\text{Earth}}}{m}\bigg)^{2}\cdot  \bigg(\frac{\text{R}_\mathrm{Earth}}{d}\bigg)^{4}.
\end{align}
Thus, if the Sun were a black hole and the Earth occupied a quantum state with its center of mass spatially superposed by a separation on the order of its own radius, this superposition would decohere due to the presence of the black hole in about $10$ $\mu$s. Of course, it would not be easy to put the Earth into such a quantum superposition.

In summary, we have found that black holes, in effect, gather information about quantum superpositions of spatially separated components by means of the long range fields sourced by the matter comprising these components. Eventually, a black hole will decohere any quantum superposition. Although this may not be of practical importance for any presently contemplated experiments, it may be of fundamental significance for our understanding of the nature of black holes in a quantum theory of gravity.

\subsection*{Acknowledgements}

D.L.D. acknowledges his support as a Fannie and John Hertz Foundation Fellow holding the Barbara Ann Canavan Fellowship, and as an Eckhardt Graduate Scholar in the Physical Sciences Division at the University of Chicago. This research was supported in part by NSF grant 21-05878 to the University of Chicago.

\bibliographystyle{JHEP.bst}
\bibliography{GRF}

\end{document}